\documentclass[12pt]{article}
\usepackage{graphicx}
\newcommand{\sss}{\scriptscriptstyle}

\newcommand {\be}{\begin{equation}} 
\newcommand{\ee}{\end{equation}}    

\def\dds1{\frac{\partial}{\partial s_1}}

\def\vti{v_{{\sss T}i}}

\def\vta{v_{{\sss T}a}}
\def\vtb{v_{{\sss T}b}}

\def\d{d\kern-0.8 ex\vrule height 1.3 ex depth-1.24 ex width 0.7 ex
\kern 0.15 ex}
\def\D{D\kern-1.7 ex\vrule height .87 ex depth-0.8 ex width 0.7 ex
\kern 0.95 ex}

\begin{document}
\baselineskip 20 pt

\begin{center}

\Large{\bf Collisional energy transfer in  two-component   plasmas }

\end{center}

\vspace{0.7cm}

\begin{center}

 J. Vranjes$^1$, M. Kono$^2$, S. Poedts$^1$, M. Y. Tanaka$^3$

{\em $^1$Center for Plasma Astrophysics, Celestijnenlaan 200B, 3001 Leuven,
 Belgium.}

{\em $^2$Faculty of Policy Studies, Chuo University, Hachioji, Tokyo 192-0393, Japan.}

{\em  $^3$Interdisciplinary Graduate School of Engineering Sciences, \\Kyushu
 University, Kasuga-koen 6-1, Kasuga, Fukuoka 816-8580, Japan.}

\end{center}

\vspace{2cm}

{\bf Abstract:} The friction in  plasmas  consisting of two species with different temperatures is discussed together with the consequent energy transfer. It is shown that the friction between the two  species  has  no effect  on the ion acoustic mode in a quasi-neutral plasma. Using the Poisson equation instead of the quasi-neutrality reveals the possibility for an instability driven by the collisional energy transfer. However, the different starting temperatures of the two species imply an evolving equilibrium. It is shown  that the relaxation time of the equilibrium electron-ion plasma is, in fact, always shorter than the growth rate time, and the instability can thus never effectively take place. The results obtained here should contribute to the definite clarification of some contradictory results obtained in the past.


\vfill

\pagebreak


Plasmas both in the laboratory and in space are frequently in the state of partial thermodynamic equilibrium \cite{ok}-\cite{mer} (i.e.\ with an initial temperature disparity of the plasma constituents). Collisions in such plasmas will after some time eventually result in equal temperatures of the species, implying an evolving  plasma. However, there exists a long standing controversy in the literature, which deals with the effects of this temperature disparity on the ion acoustic (IA) waves.

In Ref.~\cite{rs} it is claimed that the corresponding energy transfer may result in the instability of the acoustic mode at large wavelengths (within the quasi-neutrality limit), and that this growth may be described within the fluid theory. The {\em necessary} condition  for the instability obtained in Ref.~\cite{rs} for an electron-ion plasma is, in fact, very easily satisfied because it requires only a very small temperature difference between the two species (electrons and ions), viz.\  $ T_e>4 T_i/3$.
This instability condition is obtained by using the energy equations including the source/sink  terms originating from the collisional transfer, together with the corresponding friction force terms in the momentum equations. The {\em sufficient} instability condition is stronger because of additional dissipative effects, like viscosity and  thermal conductivity.

However, the current-less instability described in Ref.~\cite{rs} is based on a model which disregards the same temperature disparity in the description of the equilibrium, which, due to the same reasons, must be time evolving. In other words, the effects of collisions in the equilibrium have been explicitly neglected.    These effects  have been discussed in Ref.~\cite{bat}, published one year after Ref.~\cite{rs} and for the same quasi-neutrality case. There, it  is  claimed  that there is no instability  for any temperature ratio of the two plasma components, and moreover,   that this holds even in a current-carrying plasma, as long as the difference between the electron and ion equilibrium velocity remains below the sound speed. All that  was  needed to come to that conclusion was to let the equilibrium plasma evolve freely in the presence of the given temperature difference. However, we observe that  Ref.~\cite{bat} has apparently  remained almost  unnoticed by researchers, in contrast to the widely cited Ref.~\cite{rs}, see e.g.\ Refs.~\cite{ong}-\cite{sing} and many others.

In the present work,  this controversy is revisited for any two-component plasma.
Essential for the problem is the energy equation describing the temperature variation. In the simplified form that  we shall use, it contains only the collisional energy transfer source/sink term on the right-hand side. This  simplified form is used for clarity only because, according to Ref.~\cite{rs},  in the absence of currents, that term alone is supposed to yield an instability. In view of the controversy mentioned above, here we give some details following Braginskii \cite{brag}, where  the energy equation for any species $a$  is given in the form:
\be
\frac{3}{2} n_a\frac{\partial   T_a}{\partial t} + n_a  T_a \nabla\cdot \vec v_a + \frac{3}{2}  n_a (\vec v_a\cdot \nabla)T_a=
 Q_a. \label{en2}
\ee
The corresponding equation for the species $b$ has the same shape, but with a minus sign  on the right-hand side. We use the Landau formula for  the energy transfer source/sink term \cite{lan}, $Q_a =3m_b \nu_{ba} n_b  (T_b- T_a)/m_a$, where \cite{spit}
 \be
 \nu_{ba}= 4\left(\frac{2\pi}{m_b}\right)^{1/2} \left(\frac{q_aq_b}{4 \pi \varepsilon_0}\right)^2 \frac{n_a L_{ba}}{3(T_b +  T_a m_b/m_a)^{3/2}}.
   \label{cf}
  \ee
 The Coulomb logarithm is given by $L_{ba}=\log[r_d/b_0]$, $r_d=r_{da} r_{db}/(r_{da}^2 + r_{db}^2)^{1/2}$,  $r_{dj}= v_{{\sss T} j}/\omega_{pj}$, and  $b_0=[|q_a q_b|/(4 \pi\varepsilon_0)]/[3(T_a+ T_b)]$ is the impact parameter.

The additional source/sink term of the form $\vec F_{fa}\cdot (\vec v_a-\vec v_b)$, where $\vec F_{fa}$ is the friction force acting on the species $a$, in the absence of equilibrium currents/drifts, is in fact nonlinear and will not be discussed here.
The other (sink) terms, due to viscosity and thermal conductivity, are omitted only for the sake of clarity, i.e.\ in order to demonstrate more clearly the effect of the disputed collisional energy transfer term. The effect of these omitted terms is easily predictable. Equation~(\ref{en2}) is valid for any species $a, b$, thus including  the electron-ion plasma from Refs.~\cite{rs}-\cite{wag}.


a)~In Ref.~\cite{rs} the collisions in the equilibrium were explicitly ignored. In that case, the two  energy equations  without the equilibrium effects, corresponding to the model from Ref.~\cite{rs} read:
\[
   \frac{\partial   T_{(a,b)1}}{\partial t} + \frac{2}{3} T_{(a,b)0} \nabla\cdot \vec v_{(a,b)1}= \pm 2 \frac{m_b}{m_a}\nu_{ba}
 \left(T_{b1}- T_{a1}\right)\]
\vspace{-15pt}
\be
  \pm 2 \nu_{ba}\frac{m_b}{m_a}\left(T_{b0}- T_{a0}\right) \frac{n_{b1}}{n_0}.
\ee
Here, the minus sign applies to the species $b$.

The two  momentum equations and the two continuity equations have standard forms and there is no need to write them down here. We stress only the presence of  the friction force terms in the momentum equations. These are of the form $\vec F_{fa}=-m_a n_a\nu_{ab} (\vec v_a-\vec v_b)$ and $\vec F_{fb}=-m_b n_b\nu_{ba} (\vec v_b-\vec v_a)$, respectively.

In the case of quasi-neutral perturbations, the two number densities $n_{(a, b)1}$ are calculated from the continuity equations and are made equal assuming  a  quasi-neutral plasma,  like in Refs.~\cite{rs}, \cite{bat} (this is typically done when  dealing with  wavelengths that are much longer than the Debye length). The  dispersion equation reads:
\be
\left(\omega + \frac{i 4 m_b\nu_{ba}}{m_a}\right) \left(\omega^2 - \frac{5}{3} k^2 \frac{T_{a0} + T_{b0}}{m_a+ m_b}\right)=0.\label{en5}
\ee
Hence, even using the same model as in Ref.~\cite{rs}, we conclude that  there is neither  an instability nor damping of the acoustic mode, regardless of the ratio $T_{a0}/T_{b0}$.

Note that  the momentum conservation  condition $ \nu_{ab}=m_b n_b\nu_{ba}/(m_a n_a)$ is  {\em nowhere} used  in the derivation of Eq.~(\ref{en5}). This is because the friction terms vanish in any case. In fact,  from the two continuity equations we have  the velocities $v_{j1}=\omega n_{j1}/(k n_0)$. These expressions, together with the assumption of quasi-neutrality, cancel the friction completely. This remains so for any two species $a$ and $b$ as long as their charge numbers $Z_a$ and $Z_b$ are constant. Further in the text we assume singly charged species.


b)~The derivations are now repeated for  isothermal quasi-neutral perturbations. In addition, the energy equation may be omitted in the equilibrium also assuming that the relaxation time for the equilibrium temperature is much longer that the period of wave oscillations. Keeping the full friction force $\vec F_f$ in both momentum equations, and within  the same quasi-neutrality limit,  yields a {\em real} dispersion equation $\omega^2=k^2 (T_{a0}+ T_{b0})/(m_a+ m_b)$.  In the given limit the collisions (through friction)  do not affect the isothermal  ion acoustic mode. This fact is usually overlooked in the literature.
The collisions appear in Eq.~(\ref{en5}) only from the energy equations, yet they do not affect the IA mode.


b.1)~Using the Poisson equation instead of quasi-neutrality,  for isothermal perturbations we obtain  coupled and damped IA  and  Langmuir  waves
\[
\omega^4 + i(\nu_{ab}+ \nu_{ba})\omega^3 - \left[k^2 \left(\vta^2+ \vtb^2\right) +\omega_{pa}^2+ \omega_{pb}^2\right]\omega^2
\]
\[
- i k^2\left(\nu_{ab}\vtb^2 + \nu_{ba} \vta^2\right)\omega
\]
\vspace{-19pt}
\be
+ k^4 \vta^2 \vtb^2 + k^2 \left(\vta^2 \omega_{pb}^2+ \vtb^2 \omega_{pa}^2\right)=0. \label{iso}
\ee
In the collision-less limit  the  two modes (\ref{iso}) decouple by setting $T_a=T_b$,  though strictly speaking  this has not much sense because in this case the acoustic mode may lose its electrostatic nature, especially in pair-plasmas. For a pair (pair-ion, electron-positron)  collision-less plasma the solutions are
$\omega^2= \omega_p^2 + k^2(\vta^2+\vtb^2)/2  \pm [\omega_p^4+ k^4(\vta^2-\vtb^2)/4]^{1/2}$.

In the low frequency limit $\omega\ll \omega_{p(a,b)}$ and for an  e-i plasma,  from Eq.~(\ref{iso})  we have $\omega^2=k^2 v_s^2 - i 2 \nu_{ei}\omega m_e r_{de}^2 k^2/m_i$, so that  the IA mode is damped
\vspace{-5pt}
\be
\omega=\pm k v_s\left(1-r_{de}^2k^2 \frac{\nu_{ie}^2r_{de}^2}{v_s^2}\right)^{1/2} - i \nu_{ie}r_{de}^2k^2. \label{puas}
\vspace{-5pt}
\ee

We have used $\nu_{ie}=m_e\nu_{ei}/m_i$,  and  $v_s^2= c_s^2+ \vti^2$.  The damping in Eq.~(\ref{puas}) is $k$-dependent.


c)~From Eq.~(\ref{en2}) it is seen that in a quasi-neutral  homogeneous equilibrium, without flows/currents,  {\em the equilibrium temperature is also evolving in time} as
\be
 \frac{\partial   T_{(a,b)0}}{\partial t} = \pm  2 \frac{m_b}{m_a}\nu_{ba}  (T_{b0}- T_{a0}). \label{en3}
\ee
 Keeping the collision frequencies constant  this gives the two temperatures  $T_{(a, b)0} = [\widehat{T}_{(a, b)0} (1+ \exp(- 4 \nu_{ab} t) ) + \widehat{T}_{(a, b)0} (1- \exp(- 4 \nu_{ab} t) )]/2$ evolving towards the common value $(\widehat{T}_{a0} + \widehat{T}_{ b0})/2$.
 On the other hand, solving (\ref{en3}) numerically with time dependent collision frequencies (\ref{cf})  gives a slightly faster relaxation for the two temperatures. To get a feeling on the time scale, this is presented in Fig.~1 by taking $n_0=10^{18}$ m$^{-3}$ and  $\widehat{T}_{a0}= 0.1$ eV, $\widehat{T}_{b0}= 3 \widehat{T}_{a0}$.

\begin{figure}
\includegraphics[height=6cm, bb=10 20 580 450, clip=]{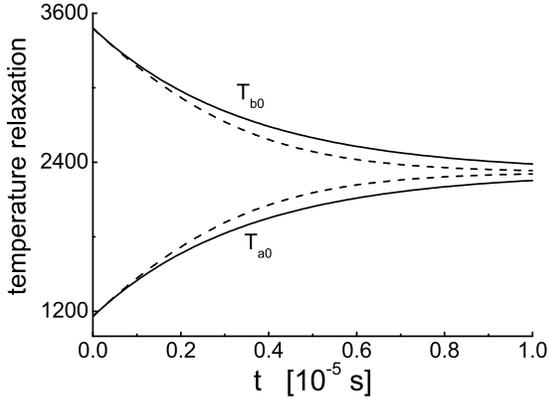}
\vspace*{-5mm} \caption{Approximative (full lines), and exact  relaxation with time-dependent collision frequencies (dashed lines) of equilibrium temperatures (\ref{en3}). }\label{fig1}
\end{figure}

Eq.~(\ref{en3}) is to be used in the linearization of Eq.~(\ref{en2}), which in the case $n_{a0}=n_{b0}=n_0$ yields:
\[
 \frac{\partial   T_{a1}}{\partial t} + \frac{2}{3} T_{a0} \nabla\cdot \vec v_{a1}= + 2 \frac{m_b}{m_a}\nu_{ba}
 \left(T_{b1}- T_{a1}\right)
 \]
 \be
    -2 \nu_{ba}\frac{m_b}{m_a} \frac{n_{a1}-n_{b1}}{n_0}\left(T_{b0}- T_{a0}\right). \label{en2a}
\ee
The corresponding equation for the component $b$ is
\be
 \frac{\partial   T_{b1}}{\partial t} + \frac{2}{3} T_{b0} \nabla\cdot \vec v_{b1}= - 2 \frac{m_b}{m_a}\nu_{ba}
 \left(T_{b1}- T_{a1}\right). \label{en2b}
\ee
Here, in the process of linearization yielding Eq.~(\ref{en2b}),  the term $(3/2) n_{b1} \partial T_{b0}/\partial t$ on the left-hand side, cancels  out with the term $- (m_b/m_a)\nu_{ba} (T_{b0}-T_{a0}) n_{b1}$ on the right-hand side after using the equilibrium equation (\ref{en3}) for the species $b$.

Hence, both Eqs.~(\ref{en2a}) and (\ref{en2b}) are obtained taking into account the evolution of the equilibrium. There appears  an additional asymmetry between the two energy equations (apart from the opposite signs of the first term on the right-hand side), due to the last term in Eq.~(\ref{en2a}). This extra asymmetry  is a consequence of the fact that the internal energy of the two species may also change  due to the presence of the new ingredient in the system, i.e., the perturbed electric field $n_{a1}-n_{b1}=\varepsilon_0 \nabla\cdot \vec E_1/e$ (in the presence of the necessary collisions of course). However, it  vanishes  if  the quasi-neutrality condition is used on the right-hand side in Eq.~(\ref{en2a}), which   sometimes may  be permissible  in  higher order terms but not in general, for example  assuming  that
the source/sink term in the energy equations gives only small imaginary corrections to  the frequency.

However, regardless of the fact that the last term in Eq.~(\ref{en2a}) is used or not, the effects of the evolving equilibrium remain within Eq.~(\ref{en2a}) in both cases.  Note also  that  the cancelation of the terms in the equation for the species $b$ (which is due to evolving equilibrium as described above) remains intact.

c.1)~We stress that  Eq.~(\ref{en5})  is  obtained also  by using Eqs.~(\ref{en2a}, \ref{en2b}) in the quasi-neutral limit (implying that the last term in Eq.~(\ref{en2a}) is  omitted). Hence, the IA mode appears unaffected by friction in the quasi-neutral limit even if the energy equations are used and the equilibrium is described correctly as evolving.

c.2)~We now use the two  energy equations (\ref{en2a}, \ref{en2b})  {\em with  the Poisson equation}.
The  dispersion equation becomes
 \[
 \omega^6 + i  \nu_{ba} \left(1+ \frac{5 m_b}{m_a}\right)\omega^5 - \omega^4 \left[\frac{5}{3} k^2 \left(\vta^2 + \vtb^2\right) + \omega_{pa}^2
  \right.
  \]
    \[
  \left.
  + \omega_{pb}^2 + 4 \nu_{ba}^2 \frac{m_b}{m_a} \left(1+ \frac{ m_b}{m_a}\right)\right] + i \omega^3 \nu_{ba} \left[  k^2 \vtb^2  \frac{2m_b^2}{m_a^2}
  \right.
  \]
    \[
  - \left. k^2 \vta^2 \left(\frac{5}{3} + \frac{22 m_b}{3 m_a}\right)\right.
  \left.
   - 4  \omega_{pa}^2 \left(1+ \frac{ m_b}{m_a}\right)  - 7 k^2 \vtb^2  \frac{m_b}{m_a}\right]
      \]
     \[
    + \omega^2 \left[ \frac{25}{9}k^4 \vta^2\vtb^2
 + \frac{5 k^2}{3} \left(\vta^2 \omega_{pb}^2 + \vtb^2 \omega_{pa}^2\right)\right.
 \]
   \[
 \left.
 + 4 k^2 \nu_{ba}^2 \frac{m_b}{m_a} \left( \vta^2\left(\frac{2}{3}+ \frac{m_b}{m_a}\right) + \vtb^2 \frac{m_b^2}{m_a^2}\left(\frac{8}{3}- \frac{m_b}{m_a}\right)\right)\right]
 \]
   \[
 + i 10 \omega \nu_{ba}  k^2\left[k^2 \vtb^2 \frac{m_b}{m_a} \left(\vta^2 - \frac{\vtb^2}{3} \frac{m_b}{m_a}\right) + \frac{2}{3}  \omega_{pa}^2  \left(\vta^2\right.\right.
 \]
   \be
 \left. \left.  + \vtb^2 \frac{m_b}{m_a}\right)\right]
   + 4 k^4 \nu_{ba}^2 \frac{m_b}{m_a} \left(\vta^2 -  \vtb^2 \frac{m_b}{m_a}\right)^2=0.  \label{de}
  \ee
Solving for the IA mode yields approximately the frequency of the IA mode
\be
\omega_{{\sss IA}}^2= \frac{5 k^2 (\omega_{pb}^2 \vta^2 +
\omega_{pa}^2 \vtb^2 + 5 k^2 \vta^2 \vtb^2/3)}{3 \left[\omega_{pa}^2 +\omega_{pb}^2  + (5/3) k^2 (\vta^2 + \vtb^2)\right]
}.
\label{ia}
\ee
The growth rate is:
\[
\gamma=\frac{1}{2 \left[ \omega_{pa}^2 + \omega_{pb}^2 + (5/3) k^2 (\vta^2 + \vtb^2) - 2 \omega_{{\sss IA}}^2\right]}\times
\]
\[
\times \left\{\nu_{ab}\left(\omega_{{\sss IA}}^2 - (5/3) k^2 \vtb^2\right)\left(1+ (4/3) k^2 \vta^2/\omega_{{\sss IA}}^2\right)\right.
\]
\[
\left. + \nu_{ba} \left(\omega_{{\sss IA}}^2 - (5/3) k^2 \vta^2\right) \left[1+ (4/3) \nu_{ab}k^2 \vtb^2/(\nu_{ba}\omega_{{\sss IA}}^2)\right]\right.
\]
\[
\left.  + 2 (1- T_{a0}/T_{b0}) \left(\nu_{ab} k^2 c_s^2/ \omega_{{\sss IA}}^2\right) \left[ \omega_{{\sss IA}}^2 - (5/3) k^2 \vtb^2 \right.   \right.
\]
\[
\left.\left. - (8 \nu_{ab}\nu_{ba}/\omega_{{\sss IA}}^2) \left(\omega_{{\sss IA}}^2- k^2 \vta^2\right)
\right.\right.
\]
\be
\left.
\left.
 +
(8 \nu_{ab}^2/\omega_{{\sss IA}}^2) \left(\omega_{{\sss IA}}^2- k^2 \vtb^2\right)\right]\right\}.\label{f}
\ee
In principle, Eq.~(\ref{f}) reveals the possibility for a growing IA mode if the Poisson equation is used instead of the quasi-neutrality in a time-evolving plasma.  For example, this can be easily demonstrated in the limit of negligible  terms originating from the last term in Eq.~(\ref{en2a}), i.e., on condition $|(T_{a1}-T_{b1})/(n_{a1}- n_{b1})|\gg |T_{a0}- T_{b0}|/n_0$, or in an alternative form,
$|(T_{a1}- T_{b1})/(T_{a0}- T_{b0})| \gg r_{db}^2 k^2 |q_b \phi_1|/T_{b0}$.
In that limit, the numerical solution of Eq.~(\ref{de})  yields the growth-rate of the  IA mode in an  electron-ion plasma  that is presented  in Fig.~2. Here,   $n_0=10^{18}$ m$^{-3}$ and we take several values of $T_e/T_i$, where $T_i=0.1\;$eV. The growth rate increases with  $T_e/T_i$ but only up to $T_e/T_i\simeq 3$. For even higher values of the temperature ratio the instability ceases, this is represented by the dashed  ($T_e/T_i\simeq 10$) line.
\begin{figure}
\includegraphics[height=6cm, bb=10 20 580 450, clip=]{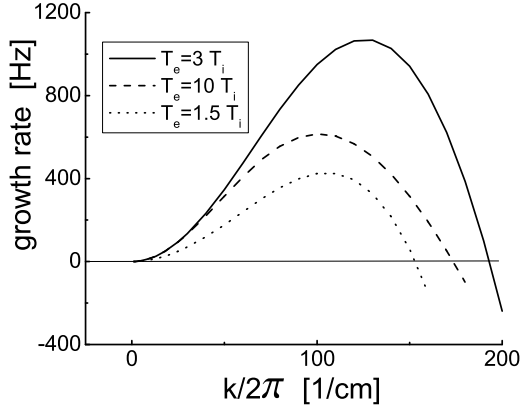}
\vspace*{-5mm} \caption{ The growth rate of the  IA mode in  electron-proton plasma with  $n_0=10^{18}$ m$^{-3}$ and for several values of $T_e/T_i$. }\label{fig2}
\end{figure}

However, we stress that the system evolves in time,  and in order to have a reasonable fast growth of the perturbations,   the following condition must be satisfied [cf.\ Eq.~(\ref{en3})] :
\be
\gamma_r\equiv 2 (m_b/m_a)\nu_{ba} \ll  \gamma.
\label{ss}
\ee
Taking the electron-ion case like  in Ref.~\cite{rs} and the corresponding self-evident conditions $m_b\ll m_a$, $T_{a0}< T_{b0}$, $k^2 \vta^2< \omega_{{\sss IA}}^2 < \omega_{pa}^2<\omega_{pb}^2$, from Eq.~(\ref{f}) to the leading order terms we obtain
\[
\gamma- \gamma_r \simeq -\frac{\nu_{ba}}{2\left(\omega_{pb}^2 + 5 k^2 \vtb^2/3\right)} \left\{
4 \omega_{pa}^2 - \omega_{{\sss IA}}^2\right.
\]
\be
\left.
+ \frac{2 k^2 c_s^2}{\omega_{{\sss IA}}^2} \left[ k^2 \vta^2 \left(\frac{5}{3} + \frac{8 \nu_{ab}^2}{\omega_{{\sss IA}}^2}\right) + 8 \nu_{ab}^2\right] \right\}. \label{f2}
\ee
Hence, because always $\omega_{pa}\geq \omega_{{\sss IA}}$, here we have
 \be
 \gamma<\gamma_r, \label{gc}
 \ee
i.e.,  the system relaxes on a time scale that is (much) shorter than the eventual growth time,  and consequently the assumed  instability actually can not  develop.
We note that  this is in agreement with some experiments, e.g.\  in  a Q-machine plasma \cite{da} where  the instability has never been observed even  by cooling the ions  to near room temperature while keeping various temperatures for the electrons.


To summarize, the long  existing controversy dealing with the stability of the ion acoustic mode in plasmas in the state of partial thermodynamic equilibrium has been revisited.  The results obtained here can be summarized as follows:  i)~The friction does not affect the IA mode in a quasi-neutral plasma; ii)~Even using the non-evolving model equivalent to Ref.~\cite{rs}, there is no instability of the IA mode, contrary to claims from Ref.~\cite{rs}; iii)~When the equilibrium plasma is properly described as evolving in time, and   as long as the quasi-neutrality is used, collisions do not produce a growht of  the ion acoustic mode;  iv)~When the Poisson equation is used instead of quasi-neutrality,  in principle there is a possibility for a positive growth-rate of the IA mode. It appears as a combined effect of  the breakdown of the charge neutrality from one side (introduced by  the Poisson equation), and  the heat transfer (the compressibility  and  advection in energy equation) from the other side, all within the background of  a time-evolving plasma. However, as the equilibrium plasma evolves in time, with the  relaxation time $\tau_r$ given in Eq.~(\ref{en3}), the obtained growth time must be (much) shorter  than the relaxation time. Yet,  this shows to be impossible  and we  conclude that there is no instability in the electron-ion plasma with an initial temperature disparity.

{\bf Acknowledgements.} Results  are  obtained in the framework of the
projects G.0304.07 (FWO-Vlaanderen), C~90205 (Prodex),  GOA/2004/01
(K.U.Leuven),  and the Interuniversity Attraction Poles Programme -
 Belgian State - Belgian Science Policy. JV  thanks N.\ D'Angelo for help and discussions.

\end{document}